\documentclass[aps,twocolumn,reprint,preprintnumbers,amsmath,amssymb]{revtex4-1}%\documentclass[twocolumn,showpacs,preprintnumbers,amsmath,amssymb]{revtex4}
\usepackage{graphicx}% Include figure files
\usepackage{dcolumn}% Align table columns on decimal point
\usepackage{bm}% bold math
\usepackage{amssymb}
\usepackage{epsfig}
\begin{document}
%\title{Extrasensitive magnetic moment detection with a proximity-nanowire quantum interferometer}
\title{Phase-controlled superconducting heat-flux quantum modulator}
%\title{Josephson heat interferometer}
\author{F. Giazotto}
\email{f.giazotto@sns.it}
\affiliation{NEST, Istituto Nanoscienze-CNR  and Scuola Normale Superiore, I-56127 Pisa, Italy}
\author{M. J. Mart\'inez-P\'erez}
\affiliation{NEST, Istituto Nanoscienze-CNR  and Scuola Normale Superiore, I-56127 Pisa, Italy}
%\date{\today}
\begin{abstract}
We theoretically put forward the concept of a phase-controlled superconducting heat-flux quantum modulator. Its operation relies on phase-dependent heat current predicted to occur in temperature-biased Josephson tunnel junctions. The device behavior is investigated as a function of temperature bias across the junctions, bath temperature, and junctions asymmetry as well. In a realistic Al-based  setup the structure could provide temperature modulation amplitudes up to $\sim 50$ mK with flux-to-temperature transfer coefficients exceeding $\sim 125$ mK$/\Phi_0$ below 1 K, and temperature modulation frequency of the order of a few MHz. The proposed structure  appears as a promising building-block for the implementation of novel-concept caloritronic devices operating at cryogenic temperatures.
 
\end{abstract}
%\pacs{72.25.-b,85.75.-d,74.50.+r}
\maketitle

The investigation of thermal transport in solid-state  nanodevices, usually referred to as \emph{caloritronics}, is a research field which is nowadays garnering an increasing attention \cite{rmp,DiVentra} since the impressive advances in nanoscience.
The latter are suggesting that caloritronics is a central issue at the nanoscale, where heat plays a key role in determining the physical properties of the system.
In this context, manipulation and control of heat currents is of great relevance for a number of applications, and  conceiving devices allowing to tune the amount of heat transferred through nanocircuits would represent an important breakthrough \cite{heatengine,Ruokola,Ojanen,brownian,modulator,segal2,casati2,linke}. Toward this direction the prototype of a thermal modulator in the form of a  heat transistor has been recently reported \cite{heattransistor,rftheo}  with the demonstration of  electrostatic control of heat currents flowing through a superconducting circuit.

Here we theoretically put forward the concept of a phase-controlled superconducting heat-flux quantum modulator.
This device exploits the phase-dependent thermal current \cite{maki} predicted to occur in temperature-biased Josephson tunnel junctions to realize heat interference.
With a realistic design, temperature modulation amplitudes up to $\sim 50$ mK with flux-to-temperature transfer functions as large as  $\sim125$ mK$/\Phi_0$  could be achieved below 1 K, and temperature modulation frequencies of the order of a few MHz. 
%This structure appears as a potential candidate for the implementation of novel-concept phase-dependent thermal devices operating at low temperature.

\begin{figure}[t!]
\includegraphics[width=\columnwidth]{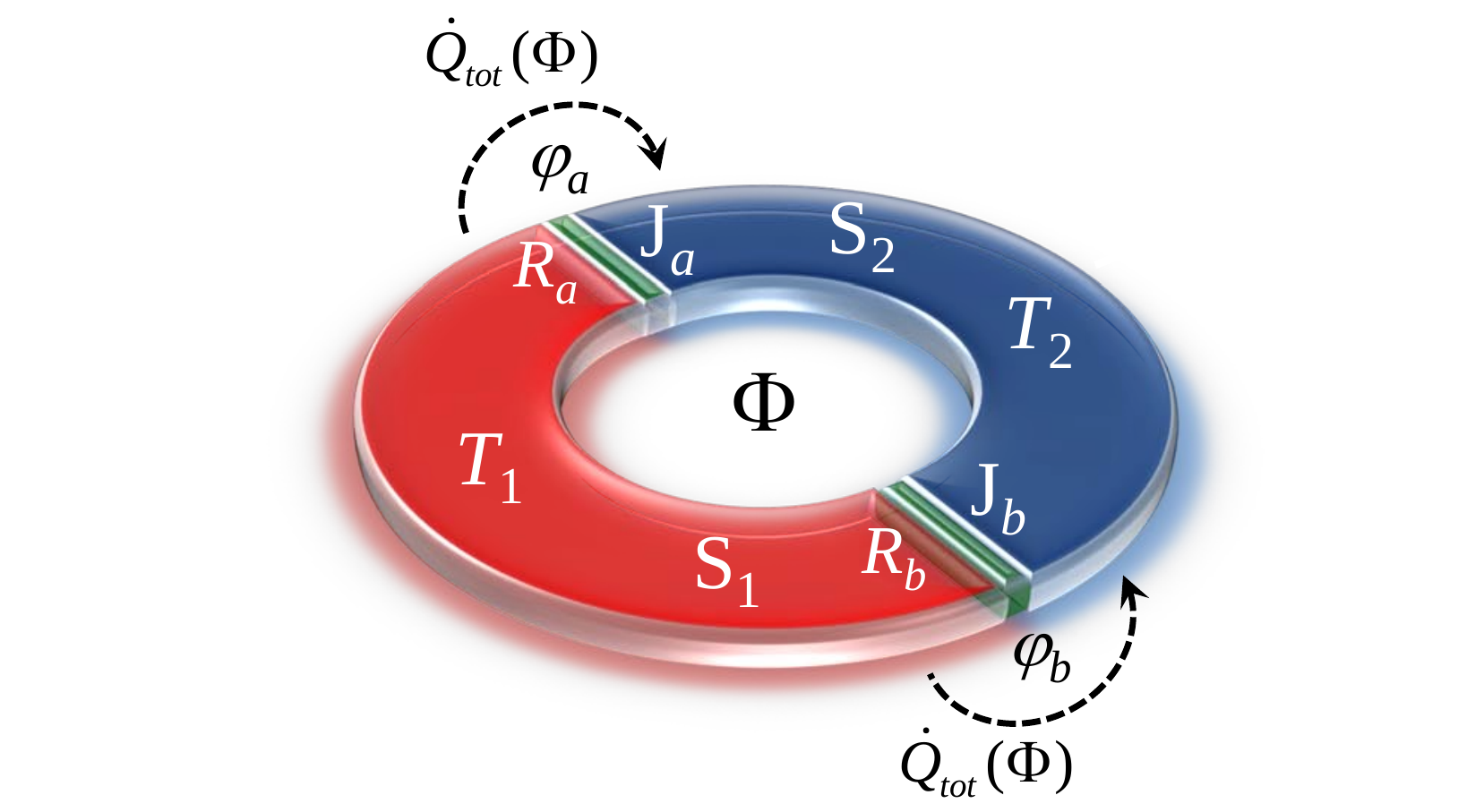}
\caption{\label{fig1} (Color online) Scheme of the proposed device.
Two superconductors S$_1$ and S$_2$ kept at temperature $T_1$ and $T_2$ (with $T_1\geq T_2$), respectively, are tunnel coupled so to implement a DC SQUID. $\varphi_{a(b)}$  is the quantum phase difference over junction J$_{a(b)}$ having normal-state resistance $R_{a(b)}$,
$\Phi$ is the applied magnetic flux threading the SQUID loop, and  $\dot{Q}_{tot}(\Phi)$ is the total heat current flowing through the structure.
}
\end{figure}

The structure we envision  is sketched in Fig. \ref{fig1}. It consists of a double-tunnel-junction superconducting quantum interference device (i.e., a DC SQUID) composed of two superconductors S$_1$ and S$_2$ in thermal equilibrium kept at different temperatures $T_1$ and $T_2$, respectively. 
$R_{a(b)}$ denotes the normal-state resistance of Josephson junction J$_{a(b)}$, while $\varphi_a$ and $\varphi_b$ represent the macroscopic quantum phase differences over the junctions. 
By neglecting the geometric inductance of the ring it follows that $\varphi_a+\varphi_b+2\pi\Phi/\Phi_0=2k\pi$ where $\Phi$ is the applied magnetic flux through the loop, $k$ is an integer and $\Phi_0=2.067\times10^{-15}$ Wb is the flux quantum.
For definiteness, we assume that $T_1\geq T_2$ so that the SQUID is only biased with a temperature drop across the junctions, but the voltage across them vanishes. 
%--------------------------------------------------figure 1------------------------------------------------------
%-------------------------------------------------------------------------------------------------------------------
In these conditions the total heat current ($\dot{Q}_{tot}$) flowing from S$_1$ to S$_2$ becomes stationary \cite{maki}, and can be written as \cite{maki,sauls1,sauls2,guttman1,guttman2} 
\begin{equation}
\dot{Q}_{tot}=\dot{Q}_{qp}(T_1,T_2)-\dot{Q}_{int}(T_1,T_2,\varphi_a,\varphi_b).
\label{qdot}
\end{equation}
In Eq. (\ref{qdot}), 
\begin{equation}
\dot{Q}_{qp}=\dot{Q}_{qp}^a(T_1,T_2)+\dot{Q}_{qp}^b(T_1,T_2)
\end{equation}
is the usual total thermal flux carried by quasiparticles through both junctions J$_{a}$ and J$_{b}$.
On the other side, 
\begin{equation}
\dot{Q}_{int}=\dot{Q}_{int}^a(T_1,T_2)\text{cos}\varphi_a+ \dot{Q}_{int}^b(T_1,T_2)\text{cos}\varphi_b,
\end{equation}
describes the sum of the heat currents due to interference between quasiparticles and Cooper pairs condensate flowing through each junction of the SQUID. We emphasize that $\dot{Q}_{int}$ originates from the Josephson effect and is peculiar to weakly-coupled superconductors \cite{maki,sauls1,sauls2,guttman1,guttman2}. According to Eqs. (\ref{qdot}) and (3), $\dot{Q}_{tot}$ therefore consists of a phase-dependent function ($\dot{Q}_{int}$) superimposed on top of a phase-independent component. 
In writing Eq. (\ref{qdot}) we have explicitly omitted the term related to heat current carried by the Cooper pairs condensate, as this term produces no dissipation in the system \cite{maki,guttman1,guttman2}.

In the aforementioned  expressions, 
$\dot{Q}_{qp}^{a(b)}=\frac{1}{e^2 R_{a(b)}}\int^{\infty}_{0} d\varepsilon \varepsilon \mathcal{N}_1 (\varepsilon,T_1)\mathcal{N}_2 (\varepsilon,T_2)[f(T_2)-f(T_1)]$ \cite{maki,sauls1,sauls2,guttman1,guttman2}, $\dot{Q}_{int}^{a(b)}=\frac{1}{e^2 R_{a(b)}}\int^{\infty}_{0}d\varepsilon \varepsilon \mathcal{M}_{1}(\varepsilon,T_1)\mathcal{M}_{2}(\varepsilon,T_2)[f(T_2)-f(T_1)]$ \cite{maki,sauls1,sauls2,guttman1,guttman2},
$\mathcal{N}_{i}(\varepsilon,T_i)=|\varepsilon|/\sqrt{\varepsilon^2-\Delta_{i}(T_i)^2}\Theta[\varepsilon^2-\Delta_{i}(T_i)^2]$ is the normalized BCS density of states in S$_{_i}$ at temperature $T_i$ ($i=1,2$),
$\mathcal{M}_{i}(\varepsilon,T_i)=\Delta_{i}(T_i)/\sqrt{\varepsilon^2-\Delta_{i}(T_i)^2}\Theta[\varepsilon^2-\Delta_{i}(T_i)^2]$,
and $\varepsilon$ is the energy measured from the condensate chemical potential.
Furthermore,
$\Delta_i(T_i)$ is the temperature-dependent energy gap, $f_i(T_i)=\text{tanh}(\varepsilon/2 k_BT_i)$,
$\Theta(x)$ is the Heaviside step function,
$k_B$ is the Boltzmann constant, and $e$ is the electron charge. 
We note that both $\dot{Q}_{qp}^{a(b)}$ and $\dot{Q}_{int}^{a(b)}$ vanish for $T_1= T_2$ \cite{maki,sauls1,sauls2,guttman1,guttman2}, while $\dot{Q}_{int}^{a(b)}$ also vanishes when at least one of the superconductors is in the normal state, i.e., $\Delta_i(T_i)=0$.

For a given $\Phi$, the phases $\varphi_a$ and $\varphi_b$
are determined by the equation $I_J^a \text{sin}\varphi_a=I_J^b \text{sin}\varphi_b$ which describes conservation of the supercurrent circulating along the loop, where
$I_J^{a(b)}\propto R_{a(b)}^{-1}$ is the Ambegaokar-Baratoff critical current \cite{Clarke}  of junction J$_{a(b)}$. By defining $r=I_J^a/I_J^b=\dot{Q}_{int}^a/\dot{Q}_{int}^b=\dot{Q}_{qp}^a/\dot{Q}_{qp}^b=R_b/R_a$ (with $0\leq r\leq 1$) as the degree of asymmetry of the SQUID junctions one gets \cite{bo}
\begin{eqnarray}
& & \text{cos}\varphi_a=\frac{r+\text{cos}(2\pi x)}{\sqrt{1+r^2+2r\text{cos}(2\pi x)}},\\
& & \text{cos}\varphi_b=\frac{1+r\text{cos}(2\pi x)}{\sqrt{1+r^2+2r\text{cos}(2\pi x)}},
\end{eqnarray}
where $x=\Phi/\Phi_0$.
With the aid of Eqs. (4) and (5) $\dot{Q}_{int}$ can be rewritten as
\begin{equation}
\dot{Q}_{int}=\dot{Q}_{int}^b(T_1,T_2)\sqrt{1+r^2+2r\text{cos}\left(\frac{2\pi\Phi}{\Phi_0}\right)},
\label{totalQ}
\end{equation} 
which is analogous to the expression for the total Josephson critical current in a DC SQUID with generic junctions asymmetry \cite{Clarke}. 
In particular, for  a symmetric SQUID ($r=1$) we get
\begin{equation}
\dot{Q}_{int}=2\dot{Q}_{int}^b(T_1,T_2)\left|\text{cos}\left(\frac{\pi\Phi}{\Phi_0}\right)\right|.
\label{symm}
\end{equation}
In the following we assume $\Delta_1(0)=\Delta_2(0)=\Delta$, with $T_c=(1.764k_B)^{-1}\Delta$ representing the common critical temperature of the superconductors.
%-------------------------------------------figure 2---------------------------------------------------------
\begin{figure}[t!]
\includegraphics[width=\columnwidth]{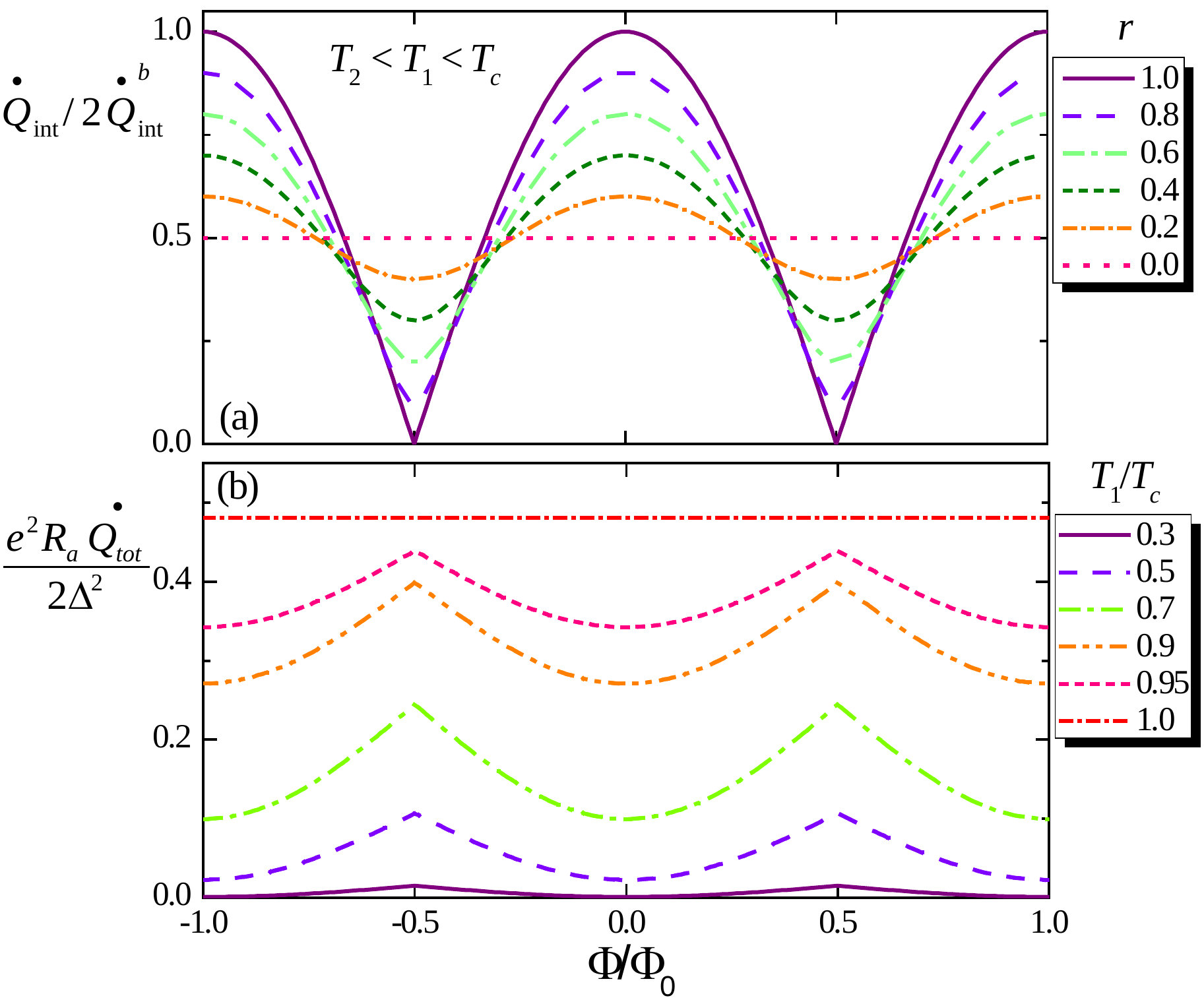}
\caption{\label{fig2} (Color online) (a) Interference heat current $\dot{Q}_{int}$ vs $\Phi$ calculated for a few values of $r$. Here we set generic temperatures $T_1$ and $T_2$ such that $T_2<T_1<T_c$.
(b) Total heat current $\dot{Q}_{tot}$ vs $\Phi$ calculated for a few values of $T_1$ at $T_2=0.1T_c$ assuming $r=1$.
}
\end{figure}
%----------------------------------------------------------------------------------------------------------------

Figure \ref{fig2} (a) shows the interference component of the heat current $\dot{Q}_{int}$ vs $\Phi$ calculated for a few values of $r$ at generic temperatures $T_1$ and $T_2$ such that $T_2<T_1<T_c$. 
As it can be seen, $\dot{Q}_{int}$ is a periodic function of $\Phi$ which is maximized at integer values of $\Phi_0$, and is modulated between the maximum given by $\dot{Q}_{int}^b(1+r)$ and the minimum given by $\dot{Q}_{int}^b(1-r)$.
For $r=1$, $\dot{Q}_{int}(\Phi)$ is thus modulated between $2\dot{Q}_{int}^b$ and 0 [see Eq. (\ref{symm})]. 
By increasing $r$ leads to a suppression of the modulation amplitude combined with a reduction of the average value of the heat current. Eventually, the modulation amplitude is totally suppressed for $r=0$, as only one junction is driving heat flow through the SQUID. 
Therefore high junctions symmetry is desired to maximize heat current modulation in the device.
For that reason, in the following we will restrict our calculations to the case $r=1$.

Figure \ref{fig2}(b) shows the total heat current $\dot{Q}_{tot}$ vs $\Phi$ calculated at $T_2=0.1T_c$ for a few values of $T_1$. 
As expected, $\dot{Q}_{tot}$ is $\Phi_0$-periodic and is minimized for integer values of $\Phi_0$ [note the sign minus in front of phase-dependent term in Eq. (\ref{qdot})]. At the lowest $T_1$ the total heat current is small, as $\dot{Q}_{qp}$ and $\dot{Q}_{int}$ are comparable in magnitude.
By increasing $T_1$ leads to the enhancement of both the average heat current and modulation amplitude which originate from larger temperature drop across the junctions. Further enhancement of $T_1$ leads to a suppression of the heat current modulation amplitude which disappears at $T_1=T_c$ when S$_1$ is driven into the normal state.
%-------------------------------------------figure 3---------------------------------------------------------
\begin{figure}[t!]
\includegraphics[width=\columnwidth]{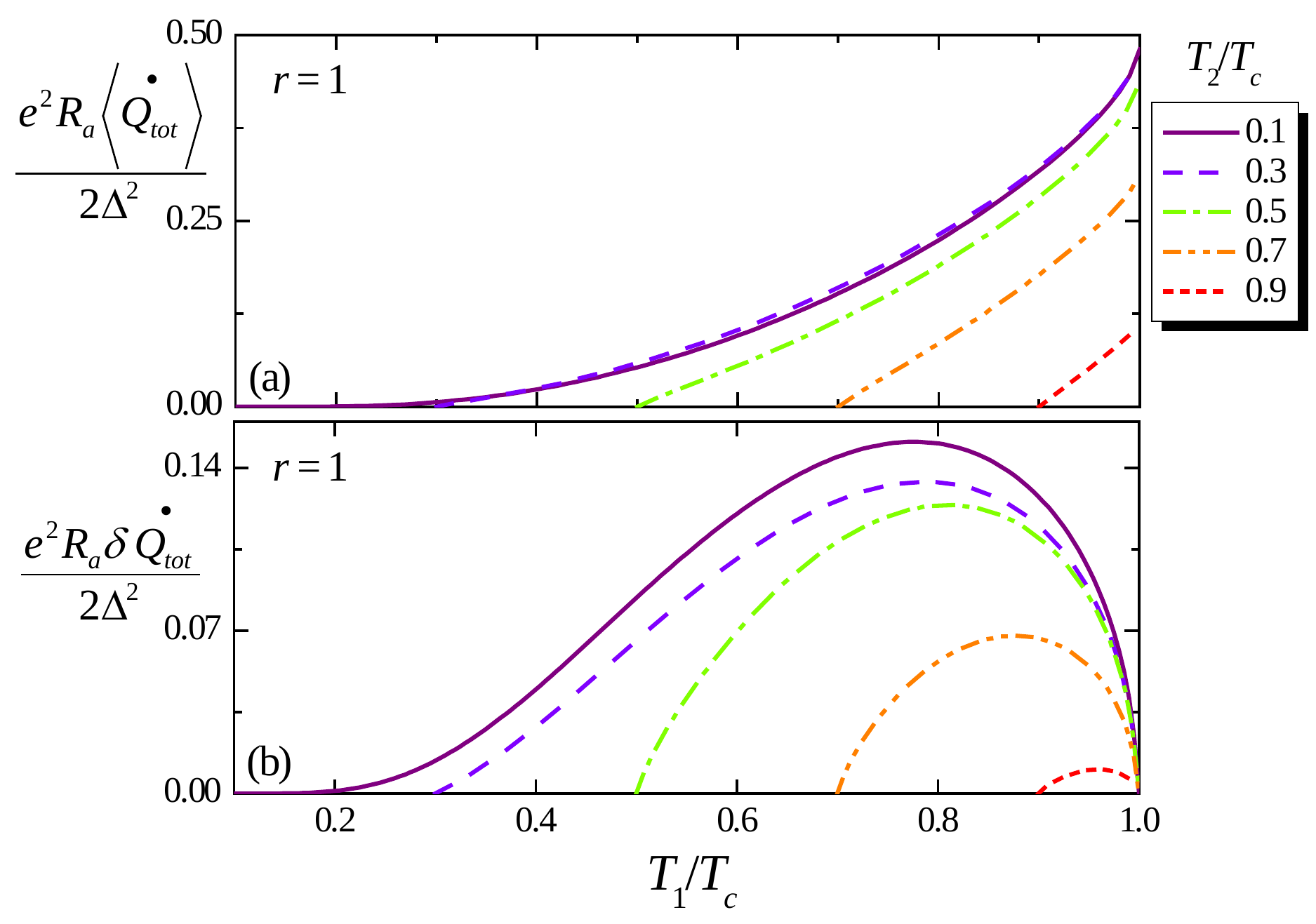}
\caption{\label{fig3} (Color online) (a) Average total heat current over one flux quantum $\left\langle \dot{Q}_{tot}\right\rangle$ vs $T_1$ calculated for some values of $T_2$. (b) Total heat current modulation amplitude $\delta \dot{Q}_{tot}$ vs $T_1$ calculated for the same $T_2$ values as in  panel (a).
}
\end{figure}
%----------------------------------------------------------------------------------------------------------------

The above described behavior is emphasized in Fig. 3. In particular,
the average $\dot{Q}_{tot}$ over one flux quantum, defined as $\left\langle \dot{Q}_{tot}\right\rangle=2(\dot{Q}_{qp}^b-\frac{2}{\pi}\dot{Q}_{int}^b)$, is plotted in Fig. \ref{fig3}(a) vs $T_1$ for some values of $T_2$.
For each $T_2$, $\left\langle \dot{Q}_{tot}\right\rangle$ monotonically increases by enhancing $T_1$, which stems from a larger absolute temperature bias across the SQUID. 
On the other hand, $\left\langle \dot{Q}_{tot}\right\rangle$ becomes more and more suppressed at higher $T_2$ for fixed $T_1$.
This suggests that a low $T_2$ combined with a substantial temperature bias are desirable to maximize the total  heat current flowing through the interferometer.

Figure \ref{fig3} (b) shows the modulation amplitude $\delta \dot{Q}_{tot}(T_1)$, defined as the difference between the maximum and the minimum values of $\dot{Q}_{tot}$, calculated for the same $T_2$ values as in panel (a). 
Specifically, for each $T_2$, $\delta \dot{Q}_{tot}$ is a non-monotonic function of $T_1$, initially increasing and saturating at an intermediate temperature  which depends on $T_2$. 
In particular, by increasing $T_2$, the  maximum of $\delta \dot{Q}_{tot}$ moves toward higher $T_1$ values.
Then, $\delta \dot{Q}_{tot}$ decreases at larger $T_1$ vanishing at $T_c$.
The behavior of $\dot{Q}_{tot}$ therefore stems from the balance among several factors such as the temperature drop, the operation temperature and the superconducting gaps.  

\begin{figure}[t!]
\includegraphics[width=\columnwidth]{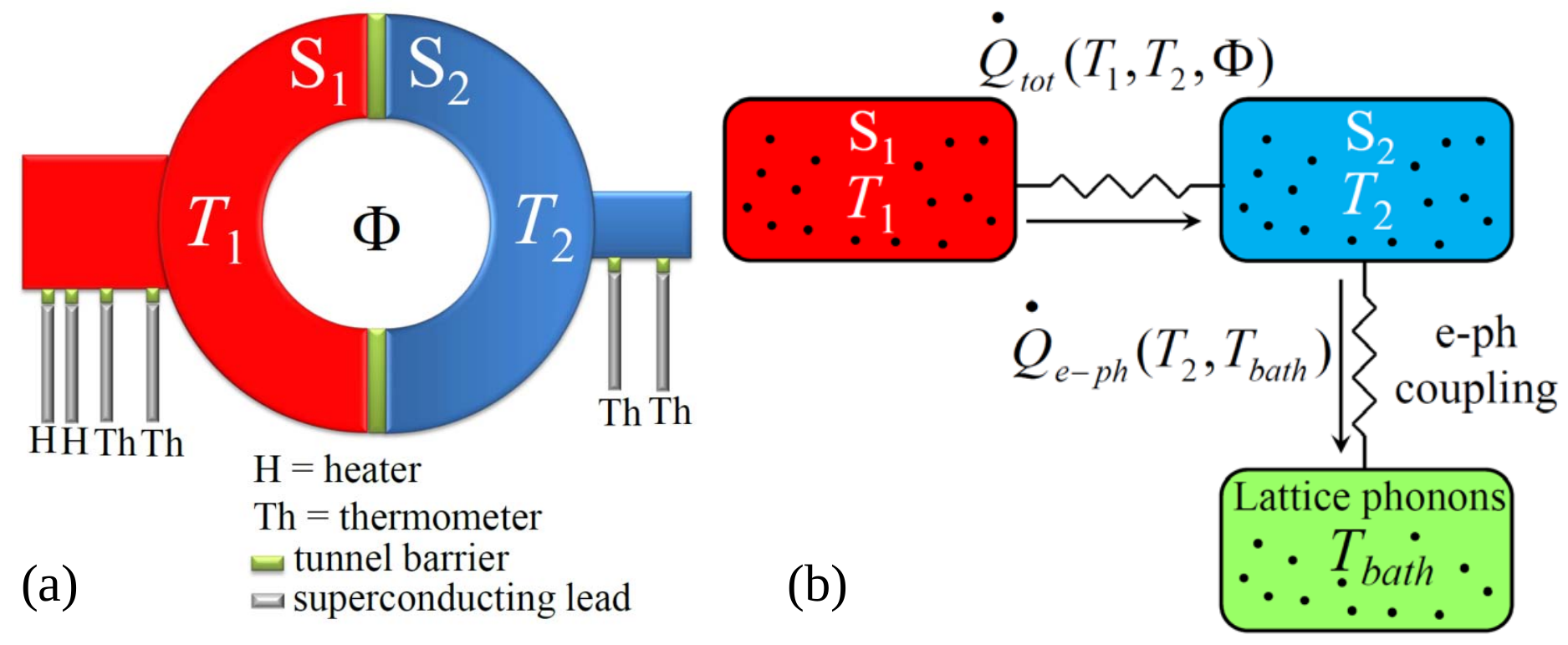}
\caption{\label{fig4} (Color online) (a) Scheme of a practical setup consisting of a temperature-biased DC SQUID threaded by a magnetic flux $\Phi$. Superconducting probes tunnel-coupled to both SQUID branches serve either as heaters (H) or thermometers (Th).
(b) Sketch of the thermal model accounting for heat transport in the system. The arrows indicate the direction of heat currents for $T_{bath}< T_2< T_1$.
}
\end{figure}

A practical experimental setup to observe these effects is shown in Fig. \ref{fig4}(a). The device can be realized with standard lithographic techniques,
and consists of a tunnel-junction DC SQUID composed of identical superconductors, and pierced by a magnetic flux $\Phi$.
Superconducting leads tunnel-coupled to both SQUID electrodes and serving either as heaters (H) or thermometers (Th) allow to perturb and to accurately probe the quasiparticle temperature in the structure \cite{rmp}.
The superconducting junctions provide nearly ideal thermal isolation of the SQUID electrodes \cite{rmp} and, therefore, we will neglect
the thermal conductance through these probes. 
%-------------------------------------------figure 4---------------------------------------------------------
%----------------------------------------------------------------------------------------------------------------

The relevant thermal model accounting for heat transport in the structure is shown in Fig. \ref{fig4}(b).  Upon  heating S$_1$ at temperature $T_1$ the steady-state quasiparticle temperature $T_2$ will depend on the energy relaxation mechanisms occurring in S$_2$. 
At low lattice temperature $T_{bath}$, typically below $\sim 1$ K, the predominant energy relaxation mechanism in metals is related to electron-phonon interaction \cite{rmp}, $\dot{Q}_{e-ph}(T_2,T_{bath})$, which allows energy exchange between quasiparticles and lattice phonons residing at $T_{bath}$. 
$\dot{Q}_{e-ph}$ in S$_2$ can be written as \cite{timofeev}
\begin{eqnarray}
& & \dot{Q}_{e-ph}=-\frac{\Sigma \mathcal{V}}{96\zeta (5)k_B^5}\int^{\infty}_{-\infty}dEE\int^{\infty}_{-\infty}d\varepsilon \varepsilon^2\text{sign}(\varepsilon)M_{E,E+\varepsilon}\nonumber\\
& & \times [\text{coth}(\frac{\varepsilon}{2k_B T_{bath}})(f_E-f_{E+\varepsilon})-f_Ef_{E+\varepsilon}+1],
\label{eph}
\end{eqnarray}
where $f_E(E,T_2)=\text{tanh}(\frac{E}{2k_B T_2})$, $M_{E,E'}(E,E',T_2)=\mathcal{N}_2(E,T_2)\mathcal{N}_2(E',T_2)[1-\frac{\Delta^2(T_2)}{EE'}]$, $\Sigma$ is the electron-phonon coupling constant, and $\mathcal{V}$ is the volume of S$_2$. 
At fixed $T_{bath}$, the steady-state $T_2(\Phi)$ is obtained by solving the  following thermal-balance equation 
\begin{equation}
-\dot{Q}_{tot}(T_1,T_2,\Phi)+\dot{Q}_{e-ph}(T_2,T_{bath})=0. 
\label{balance}
\end{equation}
In our thermal model we neglect phononic heat conduction \cite{maki}, as well as heat exchange with the photonic environment due to poor matching impedance \cite{schmidt,meschke}.
In the following calculations we assume an aluminum (Al) symmetric DC SQUID with bulk $T_c=1.19$ K, and $R_a=R_b=1$ k$\Omega$. As additional set of parameters we choose a S$_2$ electrode with volume  $\mathcal{V}=10^{-19}$ m$^3$, and $\Sigma =3\times 10^8$ Wm$^{-3}$K$^{-5}$ as typical for Al \cite{rmp}.
%-------------------------------------------figure 5---------------------------------------------------------
\begin{figure}[t!]
\includegraphics[width=\columnwidth]{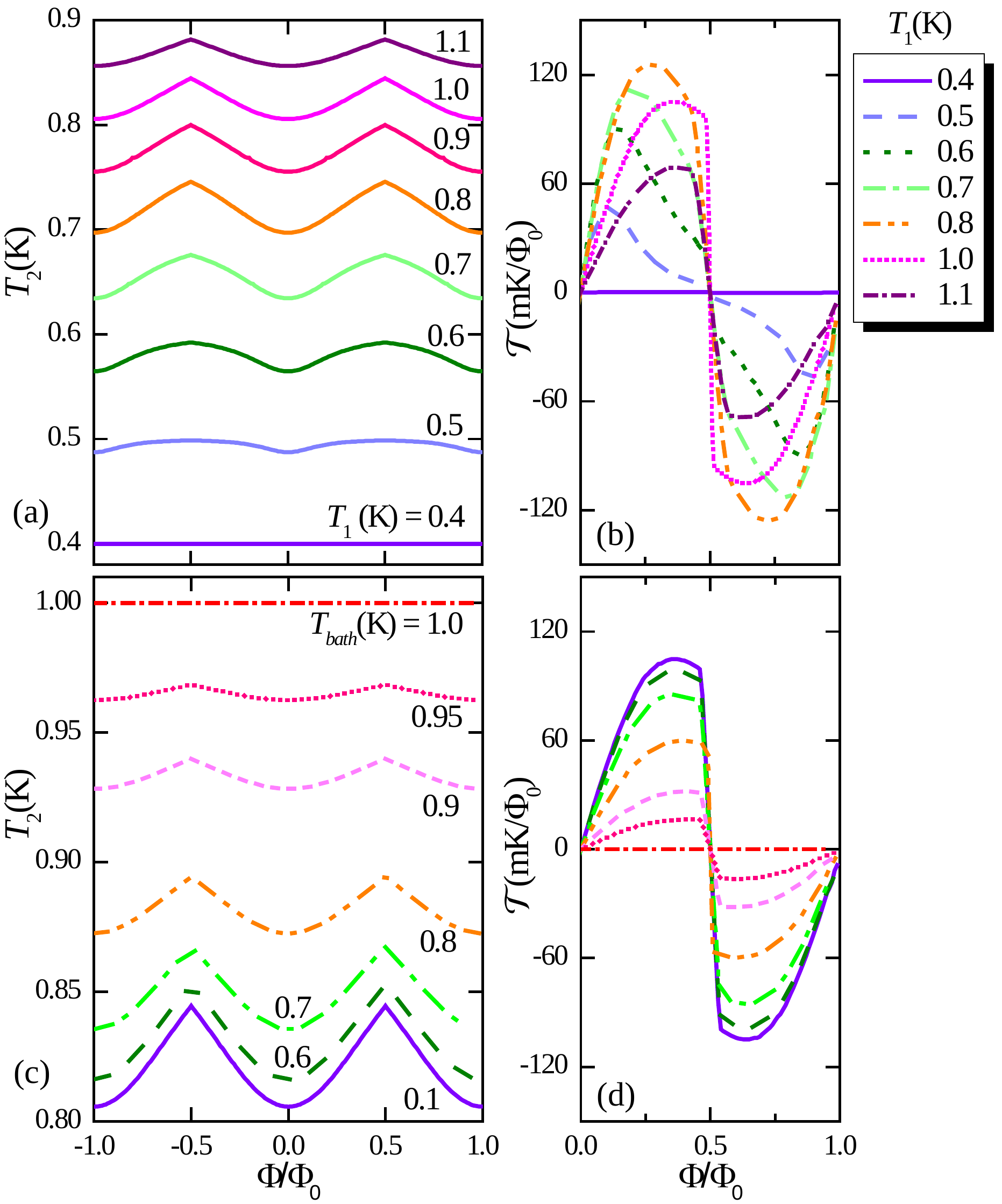}
\caption{\label{fig5} (Color online) (a) Quasiparticle temperature $T_2$ vs $\Phi$ calculated for a few values of $T_1$ at $T_{bath}=100$ mK.
(b) Flux-to-temperature transfer function $\mathcal{T}$ vs $\Phi$  calculated for selected values of $T_1$ at $T_{bath}=100$ mK.
(c) $T_2(\Phi)$ calculated for different $T_{bath}$ at $T_1=1$ K.
(d) $\mathcal{T}(\Phi)$ calculated for the same values as in panel (c).
}
\end{figure}
%----------------------------------------------------------------------------------------------------------------

The solution of Eq. (\ref{balance}) is displayed in Fig. 5(a) where $T_2(\Phi)$ is plotted for increasing $T_1$ values at $T_{bath}=100$ mK. 
In particular, by increasing $T_1$ leads to the enhancement of the average temperature $T_2$ over one flux quantum which stems from larger $\dot{Q}_{tot}$ through the interferometer.
Furthermore, while at low $T_1$ the temperature modulation amplitude $\delta T_2$ (defined as the difference between the maximum and minimum values of $T_2$) is vanishing, at higher temperature it increases obtaining up to $\sim 50$ mK at $800$ mK. At larger $T_1$, $\delta T_2$ tends to decrease due to temperature-induced suppression of the energy gaps in both superconductors. We note that this behavior  reflects basically the same picture provided by $\dot{Q}_{tot}$ [see Fig. \ref{fig2}(b)].
It is worthwhile to mention that by reducing the S$_2$ volume so to suppress $\dot{Q}_{e-ph}$ does not yield necessarily enhanced $\delta T_2$ for fixed $T_1$ and $T_{bath}$, as in such a case the temperature drop across the SQUID would decrease. 
In the opposite situation of larger  $\mathcal{V}$, enhanced $\dot{Q}_{e-ph}$ would suppress $\delta T_2$ as well. 
Analogously, by largely increasing $R_a$ leads to a reduced $\dot{Q}_{tot}$ and to a suppression of $\delta T_2$, whereas for low $R_a$ reduced temperature drop across the junctions would prevent large $\delta T_2$ values. 

In analogy to a conventional electric SQUID,  we can define for the heat-flux modulator a figure of merit  in the form of a flux-to-temperature transfer coefficient, $\mathcal{T}(\Phi)=\partial T_2/\partial \Phi$, which is plotted in Fig. 5(b) for selected values of $T_1$ at $T_{bath}=100$ mK. Specifically, $\mathcal{T}(\Phi)$ behaves similarly to $T_2(\Phi)$ as a function of $T_1$, and obtains sizable values as large as $\sim 125\,\text{mK}/\Phi_0$ at 800 mK.

The role of bath temperature is shown in Fig. 5(c) which displays $T_2(\Phi)$ calculated for increasing $T_{bath}$ values at $T_1=1$ K.
An increase in $T_{bath}$ leads to a smearing of $T_2(\Phi)$, and to a reduction of $\delta T_2$ due to the combined effect of enhanced  $\dot{Q}_{e-ph}$, reduced temperature drop and temperature-induced gap suppression in S$_2$. As expected, full suppression of $\delta T_2$ occurs at $T_{bath}=T_1$, as in such a case S$_1$ and S$_2$ are at the same temperature. 
 %up to a full suppression for $T_{bath}=T_1$, as in such a case S$_1$ and S$_2$ are at the same temperature. 
%When $T_{bath}>T_1$, $\delta T_2$ starts again to increase and obtains values as high as $\sim 45$ mK at 900 mK.
%Note that now $T_2(\Phi)$ shows negative concavity, as $T_2>T_1$. 
%At larger $T_{bath}$, $\delta T_2$  is reduced again due to the combined effect of enhanced electron-phonon interaction and temperature-induced gap suppression in S$_2$. 
This behavior directly reflects on $\mathcal{T}(\Phi)$ which is shown in Fig. \ref{fig5}(d) for the same values as in panel (c). 
%Notably, $\mathcal{T}$ as large as $\sim 130$ mK$/\Phi_0$ at 900 mK can be achieved. Moreover, $\mathcal{T}$ can be substantial even at higher $T_{bath}$ where it can obtain up to $\sim 100$ mK$/\Phi_0$ at 1.1 K.

We shall finally comment the temperature modulation speed which could be achieved in the device. The latter is mainly limited by the relaxation time $\tau$ required by quasiparticles in S$_2$ to thermalize with lattice phonons, since the  $R_aC$ time constant of the SQUID junctions ($C$ is the capacitance) can be made much smaller than $\tau$ with a suitable choice of parameters. 
In particular, in the $0.5\ldots 1$ K temperature range, $\tau^{-1}$ is of the order of $\sim 1\ldots 10$ MHz for Al \cite{kaplan}, whereas at lower $T_{bath}$ it is drastically reduced owing to increased electron-phonon relaxation time \cite{kaplan,barends}. 
We stress that the modulation frequency could be increased by more than a factor of ten by exploiting superconductors with enhanced electron-phonon coupling like, for instance, tantalum \cite{barends}.

In summary, we have proposed and analyzed a phase-controlled superconducting heat-flux quantum modulator. 
In an Al-based design, it can provide  temperature modulation amplitudes up to 50 mK and transfer functions as large as  $\sim 125$ mK$/\Phi_0$ below 1 K. This structure appears potentially attractive for the design of novel-concept thermal devices operating at low temperature.

We acknowledge F. Taddei for comments, and 
the FP7 program ``MICROKELVIN'' for partial financial support.

\end{document}